\documentclass[aps,preprint,showpacs,amsmath,superscriptaddress]{revtex4}

\usepackage{times}
\usepackage{hyperref}

\usepackage{graphicx}% Include figure files
\usepackage{dcolumn}% Align table columns on decimal point
\usepackage{bm}% bold math

\begin{document}

\title{Precision radiative corrections to the semileptonic Dalitz plot with angular correlation between polarized decaying baryon and emitted charged lepton: Effects of the four-body region
}

\author{M.\ Neri
}
\affiliation{
Escuela Superior de F{\'\i}sica y Matem\'aticas del IPN, Apartado Postal 75-702, M\'exico, D.F.\ 07738, Mexico
}

\author{A.\ Mart{\'\i}nez
}
\affiliation{
Escuela Superior de F{\'\i}sica y Matem\'aticas del IPN, Apartado Postal 75-702, M\'exico, D.F.\ 07738, Mexico
}

\author{A.\ Garc{\'\i}a
}
\affiliation{
Departamento de F{\'\i}sica, Centro de Investigaci\'on y de Estudios Avanzados del IPN, Apartado Postal 14-740, M\'exico, D.F.\ 07000, Mexico
}

\author{J.\ J.\ Torres
}
\affiliation{
Escuela Superior de C\'omputo del IPN, Apartado Postal 75-702, M\'exico, D.F.\ 07738, M\'exico
}

\author{Rub\'en Flores-Mendieta
}
\affiliation{Instituto de F{\'\i}sica, Universidad Aut\'onoma de San Luis Potos{\'\i}, \'Alvaro Obreg\'on 64, Zona Centro, San Luis Potos{\'\i}, S.L.P.\ 78000, Mexico
}

\date{\today}

\begin{abstract}
Analytical radiative corrections of order $(\alpha/\pi)(q/M_1)$ are calculated for the four-body region of the Dalitz plot of baryon semileptonic decays when the ${\hat {\mathbf s}_1} \cdot {\hat {\mathbf l}}$ correlation is present. Once the final result is available, it is possible to exhibit it in terms of the corresponding final result of the three-body region following a set of simple changes in the latter, except for a few exceptions. We cover two cases, a charged and a neutral polarized decaying baryon.
\end{abstract}


\maketitle

The radiative corrections (RC) to the decay of polarized baryons $A\to Bl\overline{\nu}_l$ require that, when the real photon in $A\to Bl\overline{\nu}_l\gamma$ cannot be detected nor kinematically discriminated, the three-body region (TBR) of the Dalitz plot be extended to the four-body region (FBR). This is the case when only the momentum of the lepton $l$ is measured. In the FBR neither the neutrino nor the photon can reach zero energy. So, this region arises solely from bremsstrahlung. In Ref.~\cite{mar00} we calculated the RC of order $(\alpha/\pi)(q/M_1)^0$ to the angular correlation ${\hat {\mathbf s}_1} \cdot {\hat {\mathbf l}}$ over the FBR. In this report we shall improve the precision of these RC by incorporating to them the contributions of order $(\alpha/\pi)(q/M_1)$. We shall follow the same notation and conventions of this reference.

One cannot easily calculate only the order $(\alpha/\pi)(q/M_1)$ RC and add them to the final results of Ref.~\cite{mar00}. It is necessary to perform the calculation from the early stages. It is a long and tedious one and, accordingly, its analytical result is long and tedious. However, once the final result is available it is possible to make a concise presentation of it by establishing a set of simple rules to convert the final result of the RC up to order $(\alpha/\pi)(q/M_1)$ over the TBR of the ${\hat {\mathbf s}_1} \cdot {\hat {\mathbf l}}$ correlation \cite{ner05} into a substantial part of the FBR result. In addition, keeping the same definitions, one can identify many of the analytical expressions in  previous works covering the unpolarized decay rate and the
${\hat {\mathbf s}_1} \cdot {\hat {\mathbf p}_2}$ correlation \cite{jua96,jua97,tor04,rfm97,tor06,tun91,mar93}. This way we can avoid many unnecessary repetitions, give a concise presentation of our results, and limit ourselves to exhibit those expressions which cannot be found in previous work.

The difference between the RC calculation in the TBR and in the FBR lies in the role of $y_0$ in the summation over the momentum of $\gamma$, where the integrations over $\varphi_k$, $y$, and $x$ are performed (let us recall that $x$ and $\varphi_k$ refer to the polar and azimuthal angles of the photon, respectively, and $y={\hat {\mathbf p}_2} \cdot {\hat {\mathbf l}}$, as defined in Ref.~\cite{tun91}). In the TBR $y_0$ appears in the integrand and as the upper limit of the $y$ integration, while in the FBR the upper limit is one and $y_0$ appears only in the integrand. Nevertheless, there is an exact parallelism between the two calculations. We shall proceed along this parallelism and introduce a subindex $F$ as a reminder of the FBR.

The decay rate over the Dalitz plot including both regions is
\begin{equation}
d\Gamma_r = d\Gamma_r^{\textrm{TBR}} + d\Gamma_r^{\textrm{FBR}},
\end{equation}
where $r=C \ [N]$ refers to a negatively charged [neutral] decaying baryon and a neutral [positively charged] emitted baryon. Other charge assignments to the baryons are discussed in Ref.~\cite{mar02}. The analytical RC to order $(\alpha/\pi)(q/M_1)$ over the TBR including the ${\hat {\mathbf s}_1} \cdot {\hat {\mathbf l}}$ correlation are found in Eq.~(1) of Ref.~\cite{ner05}. The RC over the FBR can be separated into an unpolarized part and into another one containing the ${\hat {\mathbf s}_1} \cdot {\hat {\mathbf l}}$ correlation (indicated by the upper index $s$), that is,
\begin{equation}
d\Gamma_r^{\textrm{FBR}} = d\Gamma_{rB}^{\prime \textrm{FBR}} - d\Gamma_{rB}^{(s) \, \textrm{FBR}}.
\end{equation}
The bremsstrahlung origin of these terms is indicated by the index $B$. The order $(\alpha/\pi)(q/M_1)$ RC to the spin-independent part are found for $r=C$ in Eq.~(32) of Ref.~\cite{jua96} and for $r=N$ in Eq.~(22) of Ref.~\cite{jua97}. To order $(\alpha/\pi)(q/M_1)^0$ the RC to the ${\hat {\mathbf s}_1} \cdot {\hat {\mathbf l}}$ correlation are found in Eq.~(59) of Ref.~\cite{mar00} for $r=C,N$.

We shall discuss the case $r=C$ first. Our final result with order $(\alpha/\pi)(q/M_1)$ RC to the ${\hat {\mathbf s}_1} \cdot {\hat {\mathbf l}}$ correlation is compactly given by
\begin{equation}
d\Gamma_{CB}^{(s)\, \textrm{FBR}} = \frac{\alpha}{\pi} d\Omega \, {\hat {\mathbf s}_1} \cdot {\hat {\mathbf l}} (B_2^\prime I_{C0F} + C_{AF}^{(s)}). \label{eq:dgammaC}
\end{equation}
$B_2^\prime$ is given after Eq.~(1) of Ref.~\cite{ner05}. The infrared convergent $I_{C0F}$, explicitly given in
Eq.~(37) of Ref.~\cite{mar00}, corresponds to the infrared divergent $I_{C0}$ of the TBR. $C_{AF}^{(s)}$ can be
arranged as the sum $C_{AF}^{(s)}=C_{IF}+C_{IIF}+C_{IIIF}$, where
\begin{equation}
C_{IF} = \sum_{i=1}^9 Q_{i+5} \Lambda_{iF},\quad C_{IIF} = \sum_{i=6}^{15} Q_i\Lambda_{(i+4)F},\quad \textrm{and} \quad C_{IIIF} = \sum_{i=16}^{25} Q_i \Lambda_{(i+4)F}.
\end{equation}

The $Q_i$, with $i=6,7,\ldots,25$, are quadratic functions of the form factors and are common to both regions. Their explicit expressions are found in Appendix A of Ref.~\cite{tor04}. It should be clear that for $Q_6$ and $Q_7$ we use the $\tilde{Q}_6$ and $\tilde{Q}_7$ of this Appendix, where the contributions of order $(\alpha/\pi)(q/M_1)^2$ and higher have been subtracted.

The analytical form of $C_{AF}^{(s)}$ is obtained by performing explicitly the triple integrals over the real photon variables, contained in the $\Lambda_{iF}$ functions, $i=1,2,\ldots,29$. The integration over $\varphi_{k}$ and $y$ lead to a set of functions $X_{iF}$, $Y_{iF}$, $Z_{iF}$, $I_{F}$, $\eta_{0F}$, $\gamma_{0F}$, $\chi_{ijF}$, and $\zeta_{ijF}$, which connect the $\Lambda_{iF}$ with the functions $\theta_{jF}$ that result from the third integration over $x$.

Once the final results for the FBR are available, one can see that the $\Lambda_{iF}$ and the intermediate functions can be obtained from the corresponding $\Lambda_i$ and $X_i$, $Y_i$, $Z_i$, $I$, $\eta_0$, $\gamma_0$, $\chi_{ij}$, and $\zeta_{ij}$ of the TBR by making some simple changes in the latter. This is possible because of the parallelism mentioned above. These changes are: (1) a subindex $F$ is attached to all the corresponding functions of the TBR, (2) the terms proportional to the factor $(1-y_0)$ are replaced by zero, and (3) otherwise the factor $y_0$ is kept as such. There are three exceptions to rule (3). In $I$, $\chi_{11}$ and $\eta_0$, $y_0$ appears by itself only once and there it must be replaced by $y_0=1$ to produce $I_F$, $\chi_{11F}$, and $\eta_{0F}$, respectively. The $\Lambda_i$ are found in Ref.~\cite{ner05} and the $X_i$, $Y_i$, $Z_i$, $I$, $\eta_0$, $\gamma_0$, and $\zeta_{31}$ needed here are found in Appendix B of Ref.~\cite{tor04}. Most of the $\chi_{ij}$, and $\zeta_{ij}$ are found in Sec.~IV of Ref.~\cite{rfm97}. However, these rules cannot be easily applied to $\zeta_{10}$, $\chi_{10}$, and $\chi_{20}$ because their content of $\theta_i$ functions was not made explicit. Their FBR counterparts are directly given in Ref.~\cite{tor06}. We may then limit ourselves to give explicitly $\zeta_{20}$, $\zeta_{22}$ and $\chi_{30}$ which are not found in our previous work, namely,
\begin{eqnarray*}
\zeta_{22} & = & p_2ly_0 \left[ \theta_6 - 2E(\theta_2-\theta_3) \right] - \frac12 \theta_{23} - 2l^2 \left[
E_\nu^0 \theta_2 + \frac{l-2p_2y_0}{\beta} (\theta_2-\theta_3) \right] \\
&  & \mbox{} + 2l^2\left[ \frac{3(E_\nu^0+E)}{\beta^2}(\theta_2-2\theta_3+\theta_4) - \frac{3E}{\beta^2}
(\theta_3-\theta_4-\beta \theta_5) \right],
\end{eqnarray*}
\begin{equation*}
\chi_{30} = l\eta_0\left[ p_2l(1-y_0) - 2p_2^2\right],
\end{equation*}
and
\begin{equation*}
\zeta_{20}=-p_2l(1-y_0^2).
\end{equation*}
These three functions may serve to illustrate the application of the above three rules. Applying them one obtains for the FBR
\begin{eqnarray*}
\zeta_{22F} & = & p_2ly_0\left[ \theta_{6F} - 2E (\theta_{2F}-\theta_{3F}) \right] - \frac{1}{2} \theta_{23F}
- 2l^2\left[ E_\nu^0\theta_{2F} + \frac{l-2p_2y_0}{\beta}(\theta_{2F}-\theta_{3F}) \right] \\
&  & \mbox{} + 2l^2\left[ \frac{3(E_\nu^0+E)}{\beta^2}(\theta_{2F}-2\theta_{3F}+\theta_{4F}) - \frac{3E}{\beta^2}
(\theta_{3F}-\theta_{4F}-\beta \theta_{5F}) \right],
\end{eqnarray*}
\begin{equation*}
\chi_{30F}=-2p_2^2l\eta_{0F},
\end{equation*}
and $\zeta_{20F}=0$.

To complete our analytical result, all we need now is to give the explicit expressions of the $\theta_{iF}$ functions. It is only at this point that we can make a direct use of the $\theta_{iF}$ of the order $(\alpha/\pi)(q/M_1)^0$ RC of Ref.~\cite{mar00}. For $i=2,\ldots,16$ they are listed in Appendix B of this reference. $\theta_{0F}$ is given in Eq.~(38) of this same reference. $\theta_{1F}$ does not appear in this FBR, so we can arbitrarily set it equal to zero. For $i=17,\ldots,23$ the $\theta_{iF}$ are new with respect to Ref.~\cite{mar00}. One has that $\theta_{17F}=0$ and $\theta_{18F}=1$. For $i=19,\ldots,22$ one can make a connection with the order $(\alpha/\pi)(q/M_1)$ RC to the ${\hat {\mathbf s}_1} \cdot {\hat {\mathbf p}_2}$. These $\theta_{iF}$ $(i=19,\ldots,22)$ are found in Ref.~\cite{tor06}.

Only $\theta_{23}$ and $\theta_{23F}$ remain to be identified. Their definitions are $\theta_{23}=\int_{-1}^1\xi_5(x)/(1-\beta x)^2dx$ and $\theta_{23F}=\int_{-1}^1\xi_5^T(x)/(1-\beta x)^2dx$, where $\xi_5(x)$ is found in Eq.~(36) of Ref.~\cite{tun91} and $\xi_5^T(x)$ is found in Eq.~(A6) of Ref.~\cite{jua96}. After integrating over $x$, one has $\theta_{23}=(T_{23}^+ + T_{23}^-)/p_2$ and $\theta_{23F}=T_{23F}^+ + T_{23F}^-$. The explicit expressions of the TBR $T_{23}^\pm$, namely,
\begin{eqnarray*}
T_{23}^\pm & = & 4 \left\{-3E^2 \left(\frac{\beta E_\nu^0+l-p_2}{1-\beta^2} \right) + \frac{p_2[(E_\nu^0)^2-2l^2\eta_0]}{1-\beta^2} + \frac{3E^2}{2}\left[\beta E_\nu^0I_1+2(l-p_2)(I_1-1) \right] \right. \\
&  & \mbox{} + l \left( \frac{E_\nu^0x_0^\pm}{b^\pm}\right)^2 \ln \left| \frac{a^\pm+x_0}{a^\pm \pm 1}\right| \mp \left[ \beta l(x_0^\pm)^2-b^\pm (E_\nu^0+lx_0) \right] \frac{(E_\nu^0)^2(1\mp x_0)}{b^\pm(1 \mp \beta)(1-\beta x_0)} \\
&  & \mbox{} \pm E_\nu^0 \left. \left[ 3E^2\frac{(1\mp x_0)(2\mp \beta)}{1\mp \beta} \pm \left(lE_\nu^0\left(\frac{x_0^\pm}{b^\pm}\right)^2 + 3E^2\frac{2-\beta x_0}{\beta}\right) \ln \left| \frac{1\mp \beta }{1-\beta x_0}\right| \right] \right\},
\end{eqnarray*}
become in the FBR case
\begin{eqnarray*}
T_{23F}^{\pm } & = & \frac{1}{1-\beta^2}\left\{ 8\left[\left(E_\nu^0\right)^2-5l^2+6E^2\right] + 4p_2l\beta\left( \frac{(y_0^-)^2}{b^-} - \frac{(y_0^+)^2}{b^+} \right) \right\} \\
&  & \mbox{} - 2p_2l\beta \left[ \frac{12E^2}{p_2l\beta} + \left(\frac{y_0^+}{b^+} \right)^2 - \left( \frac{y_0^-}{b^-}\right)^2 \right] I_1 \mp 4p_2l\left( \frac{y_0^\pm}{b^\pm} \right)^2I_2^\pm.
\end{eqnarray*}

In these expressions one has $x_0=-(p_2y_0+l)/E_\nu^0$, $x_0^\pm=x_0+a^\pm$, $b^\pm=1+\beta a^\pm$, $y_0^\pm = y_0\pm a^\pm$, $y_0=[(E_\nu^0)^2-p_2^2-l^2]/(2p_2l)$, and $a^\pm=(E_\nu^0\pm p_2)/l$, whereas $I_1$ and $I_2^\pm$ are found in Appendix C of Ref.~\cite{rfm97}.

Let us now proceed to our second case, $r=N$. We follow the same parallelism which allows us to start at
\begin{equation}
d\Gamma_{NB}^{(s)\, \textrm{FBR}}=\frac{\alpha}{\pi}d\Omega \, {\hat {\mathbf s}_1} \cdot {\hat {\mathbf l}} \left[ B_2^\prime I_{N0F} + C_{AF}^{(s)} + C_{NAF}^{(s)} \right],
\end{equation}
where $B_2^\prime$ and $C_{AF}^{(s)}$ are those of Eq.~($\ref{eq:dgammaC}$). The infrared-convergent function $I_{N0F}$ corresponds to the infrared-divergent $I_{N0}$ of Eq.~(35) of Ref.~\cite{mar93} and it can be found in
Ref.~\cite{tor06}.

The new summand with respect to Eq.~($\ref{eq:dgammaC})$, $C_{NAF}^{(s)}$, can be arranged as $C_{NAF}^{(s)} = D_3\rho_{N3F} + D_4\rho_{N4F}$, where $D_3=2(f_1g_1-g_1^2)$ and $D_4=2(f_1g_1+g_1^2)$. The $\rho_{N3F}$ and $\rho_{N4F}$ may be expressed as $\rho_{N3F}=\rho_{IF}+\rho_{IIF}+\rho_{IIIF}$ and $\rho_{N4F}=\rho_{IF}^\prime + \rho_{IIF}^\prime + \rho_{IIIF}^\prime$. The final result is obtained after performing the three integrations over the photon momentum. One can see from looking at this result that the analytical expressions of $\rho_{mF}$ and $\rho_{mF}^\prime$, $m=I,II,III$, can be obtained from the corresponding ones of Ref.~\cite{ner05} by applying the above three rules of the $r=C$ case. So there is no need to give the FBR expressions explicitly here. However there is a simplification worth mentioning, namely, $\rho_{IF}=\rho_{IF}^\prime=0$.

We have made crosschecks between numerical integrals and analytical results of the $\Lambda_{iF}$, $\rho_{IF}$, and $\rho_{IF}^\prime$ and they were satisfactory. With the present report we complete a program of systematically calculating model independent precision RC to the experimentally more accessible observables of baryon semileptonic decays $A\to Bl\overline{\nu}_l$ when $A$ is unpolarized and when it is polarized. In the latter case we covered the angular correlations ${\hat {\mathbf s}_1} \cdot {\hat {\mathbf l}}$ and ${\hat {\mathbf s}_1} \cdot {\hat {\mathbf p}_2}$. Our results can be used for all the six charge assignment to $A$ and $B$ expected from the light and heavy quark content of these baryons \cite{mar02}, the charged lepton $l$ may be $e^\pm$, $\mu^\pm$, or $\tau^\pm$, they are suitable for model independent experimental analysis of high statistics (several hundreds of thousands of events) hyperon decays and medium statistics (several thousands of events) of heavy quark baryons, and whether the real photon is kinematically discriminated or not. The model independence of our results originates in the Low theorem \cite{low,chew} for the bremsstrahlung photons and in the extension to all baryons \cite{gar82} of the model independent virtual RC to neutron beta decay \cite{sirlin}. Future improvements of
precision to order $(\alpha/\pi)(q/M_1)^2$ and higher will require introducing model dependent RC.

The authors acknowledge financial support from CONACYT, COFAA-IPN, and FAI-UASLP (M\'exico).

\end{document}